\documentclass[twocolumn,aps,pra,amsmath,amssymb,superscriptaddress,tightenlines]{revtex4-2}
\usepackage{amssymb}
\usepackage{amsmath}
\usepackage{graphicx}
\usepackage{color}
\usepackage{subfigure}
\usepackage{amsfonts}
\usepackage[colorlinks,urlcolor=blue,linkcolor=blue,citecolor=blue]{hyperref}

\def\narrowtext{\par\global\columnwidth20.5pc
	\global\hsize\columnwidth\global\linewidth\columnwidth
	\global\displaywidth\columnwidth}

\begin{document}
	
\title{Chirally frustrated superradiant phases in a Jaynes-Cummings trimer}	
\author{Lin-Lin Jiang}
\affiliation{Key Laboratory of Low-Dimensional Quantum Structures and Quantum Control of Ministry of Education, Key Laboratory for Matter Microstructure and Function of Hunan Province, Department of Physics and Synergetic Innovation Center for Quantum Effects and Applications, Hunan Normal University, Changsha 410081, China}
\author{Ya-Ping Ni}
\affiliation{Key Laboratory of Low-Dimensional Quantum Structures and Quantum Control of Ministry of Education, Key Laboratory for Matter Microstructure and Function of Hunan Province, Department of Physics and Synergetic Innovation Center for Quantum Effects and Applications, Hunan Normal University, Changsha 410081, China}
\author{Xuan Xie}
\affiliation{Key Laboratory of Low-Dimensional Quantum Structures and Quantum Control of Ministry of Education, Key Laboratory for Matter Microstructure and Function of Hunan Province, Department of Physics and Synergetic Innovation Center for Quantum Effects and Applications, Hunan Normal University, Changsha 410081, China}
\author{Lin Tian}
\affiliation{School of Natural Sciences, University of California, Merced, California 95343, USA}
\author{Jin-Feng Huang}
\email{Contact author: jfhuang@hunnu.edu.cn}
\affiliation{Key Laboratory of Low-Dimensional Quantum Structures and Quantum Control of Ministry of Education, Key Laboratory for Matter Microstructure and Function of Hunan Province, Department of Physics and Synergetic Innovation Center for Quantum Effects and Applications, Hunan Normal University, Changsha 410081, China}
\affiliation{Institute of Interdisciplinary Studies, Hunan Normal University, Changsha 410081, China} 
\affiliation{Hunan Research Center of the Basic Discipline for Quantum Effects and Quantum Technologies, Hunan Normal University, Changsha 410081, China}

\begin{abstract}
We investigate the emergence of frustrated quantum phases in a Jaynes-Cummings (JC) trimer with complex hopping amplitudes between the cavities, which represents the smallest frustrated unit in light-matter systems. The complex hopping amplitudes that can be engineered via synthetic gauge fields introduce chiral effects and geometric frustration to the system.
We obtain analytic solutions in the semiclassical limit and map out the phase diagram of this model, featuring one normal and three distinct superradiant phases. Among these phases, a chirally frustrated superradiant phase emerges, characterized by broken chiral and translational symmetries and unidirectional photon flow. These results reveal how frustration and symmetry breaking can arise in JC systems with synthetic gauge fields and strong light-matter coupling.
\end{abstract}
\maketitle	
\narrowtext

\section{Introduction}~\label{introduction}	
Frustration~\cite{Diep2005} is an intriguing theme in the study of many-body systems, where competing interactions prevent a system from simultaneously minimizing all local energies. It gives rise to a wealth of nontrivial phenomena, such as spin liquids~\cite{Fu2015,Balents2010}, complex ordering patterns~\cite{Bertoldi2014}, and unconventional excitations~\cite{Becca2007,Mohanta2022}. Traditionally studied in magnetic systems~\cite{Lacroix2011} and geometrically constrained lattices~\cite{Wannier1950,Moessner2006}, frustration has recently emerged as a key concept in artificial quantum platforms. The ability to engineer frustration in controllable quantum systems such as optical lattices~\cite{Gross2017}, trapped ions~\cite{Kim2010,Islam2013}, and superconducting circuits~\cite{Wang2018} opens new pathways to study exotic quantum phases and critical behavior under designated conditions. In these systems, the interplay between frustration and broken symmetry (such as time-reversal symmetry) is particularly intriguing, which can lead to directional transport~\cite{Zhao2024NP} and topological features~\cite{Drisko2017}.

Controllable light-matter interactions in cavity and circuit-QED platforms have enabled the exploration of collective quantum phenomena, ranging from the superradiant phase transition to quantum emulation of complex many-body Hamiltonians. In particular, Jaynes-Cummings (JC) lattices composed of interconnected JC models, in which each cavity interacts with a two-level system, have attracted attention because of their ability to exhibit emergent photonic phases, quantum criticality, and  strongly correlated states~\cite{Hwang2016,Tian2015,Tian2015PRB,Tian2017PRB,Tian2024PRA,Koch2009,Greentree2006, Hayward2012PRL}. Recent advances in strong light-matter coupling ~\cite{Huang2020} enable the observation of new forms of symmetry breaking and phase transitions~\cite{Liu2024} in JC lattices and the Dicke model~\cite{Dicke1954,Wang1973, Emary2003,Hepp1973,Huang2023,Schiro2012, Baumann2010,Nagy2010, Viehmann2011,Huang2009,Ciuti2010,He2024,Lieb1973,Li2006}. Furthermore, the engineering of synthetic gauge fields in photonic systems has made it possible to introduce complex hopping amplitudes between cavities, enabling the study of chiral effects and frustration~\cite{Zhang2021,Roushan2017}. In \cite{Hwang2022, Cheng2023}, superradiant phases with either frustration or chirality were studied in light-matter trimers, facilitated by counterrotating interactions~\cite{Wang1973,Emary2003,Hepp1973,Huang2023,Schiro2012,Baumann2010,Nagy2010,Viehmann2011,Hwang2022,Huang2009,Ciuti2010,He2024,Lieb1973,Li2006,Hwang2022,Cheng2023,Hwang2015,Xie2025,Ying2021,Zhang2021,Padilla2022,Xu2024,Xie2025,Liu2017,Hwang2018,Lyu2024,Chesi2024,Chesi2020,chen, cai,Zheng,Peng2024}. Meanwhile, it was shown that even without the counterrotating terms, superradiant phase transitions can occur in JC models~\cite{Huang2020,Hwang2016,Liu2024}. It is hence natural to ask whether frustration and chirality can occur in light-matter systems without counterrotating interactions. 

Here we study frustrated superradiant phases in a Jaynes-Cummings trimer, a three-site JC lattice with photon hopping between neighboring cavities. Despite its simplicity, this model represents the smallest light-matter system in which geometric frustration, spontaneous symmetry breaking, and chirality can emerge through engineered complex hopping amplitudes. Using a mean-field approach, we derive analytic solutions and map out the phase diagram of the system, which contains four distinct phases: a normal phase (NP) below a critical coupling strength and three superradiant phases (SPs) above the critical coupling. By tuning the phase of the hopping amplitudes, we identify a uniform superradiant phase (USP), in which the U(1) symmetry is spontaneously broken and a Goldstone mode emerges; a frustrated superradiant phase (FSP), in which translational symmetry is broken; and a chirally frustrated superradiant phase (CFSP), in which chiral symmetry is broken and a persistent unidirectional photon current appears. We further characterize these phases through their excitation spectra and discuss their experimental signatures. 
Recent experiments demonstrated that qubit-cavity couplings can reach the same order of magnitude as or even exceed the qubit or cavity frequencies~\cite{Huang2020,Liu2024,Crisp1991,Niemczyk2010,Todorov2010,Yoshihara2017}. 
The experimentally accessible parameters can approach the regime considered in our model. The study in this work hence provides a route toward realizing frustrated and chiral superradiant phases in controllable quantum devices. Our results reveal how synthetic gauge fields, frustration, and strong light-matter interactions can combine to generate unconventional collective phases, opening opportunities for studying symmetry breaking and nonequilibrium photon transport in few-body quantum systems.

Our work complements the quantum Rabi triangle studied in Ref.~\cite{Zhang2021}, in which frustration and chirality arise from the interplay between synthetic gauge fields and counterrotating light-matter interactions and the superradiant phase transition breaks the discrete $Z_2$ symmetry of the quantum Rabi model. In contrast, the present work demonstrates that frustrated and chiral superradiant phases can emerge in a Jaynes-Cummings trimer without counterrotating interactions, where the transition to the superradiant phase is associated with the spontaneous breaking of a continuous U(1) symmetry. Furthermore, to elucidate the connection between these two models, we investigate the evolution from the JC trimer to the quantum Rabi triangle by introducing anisotropic counterrotating interactions. The resulting critical coupling for the NP-SP transitions is presented in the Appendix. 

The remainder of this paper is organized as follows. In Sec.~II, we introduce the Jaynes-Cummings trimer model and present its phase diagram. In Sec.~III, we analyze the normal phase and derive the critical coupling for the onset of superradiance. In Sec.~IV, we investigate the superradiant phases and obtain their excitation spectra. In Sec.~V, we characterize the frustrated and chirally frustrated superradiant phases through the cavity displacement order parameters and discuss their critical behavior. In Sec.~VI, we study the photon current and chirality associated with the broken chiral symmetry. Finally, in Sec.~VII, we discuss experimental feasibility and the effects of dissipation, and conclude with a summary of the main results.

\begin{figure}[tbp]
\center
\includegraphics[clip, width=8cm]{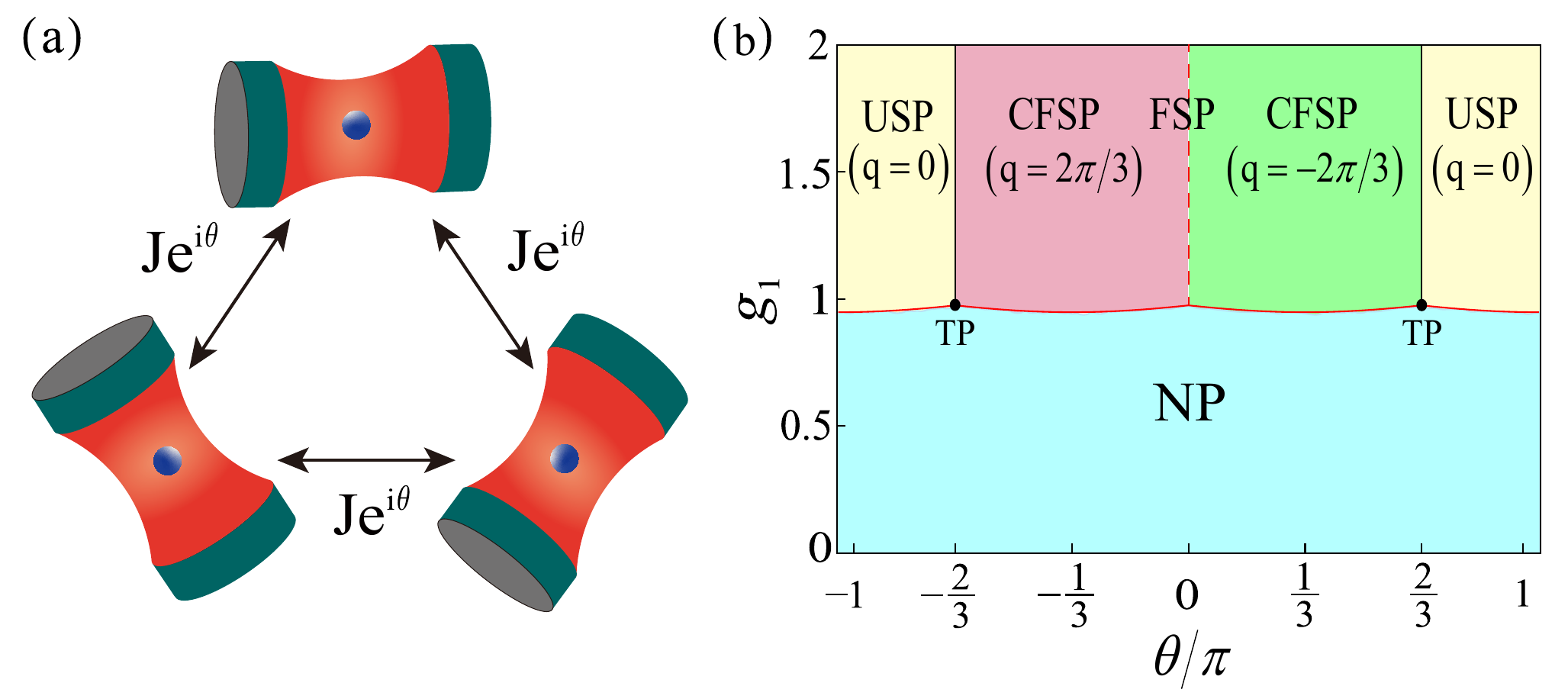} 
\caption{(a) Schematic of the JC trimer. The JC models are connected by photon hopping between neighboring sites with complex hopping amplitude $Je^{i\theta}$. (b) Phase diagram of the JC trimer for $J/\omega_{c}=0.05$.TP:Triple point}
\label{Fig1_model}
\end{figure}
\section{Model}
Consider a JC trimer, in which three JC models are connected in triangle as illustrated in Fig.~\ref{Fig1_model}(a). Each JC model is composed of a cavity mode coupled to a two-level atom, and cavity photons can hop between neighboring sites. 
The total Hamiltonian has the form: $H=\sum_n [H_{_\mathrm{JC}}^{(n)}+ H_{\mathrm{hop}}^{(n)}]$, which includes the Hamiltonian for the JC model on site $n$
\begin{equation}
    H_{_\mathrm{JC}}^{(n)} = \frac{\omega_{0}}{2}\sigma^{z}_{n}+\omega_{c}a_{n}^{\dagger}a_{n}+g(a_{n}^{\dagger}\sigma^{-}_{n}+\sigma^{+}_{n}a_{n}),
\end{equation}
and the Hamiltonian describing the photon hopping
\begin{equation}
    H_{\mathrm{hop}}^{(n)} = Je^{i\theta}a_{n}^{\dagger}a_{n+1}+\mathrm{H.c.}
\end{equation}
Here $\omega_{0}$ ($\omega_{c}$) is the frequency for the two-level atoms (cavity modes), $g$ is the strength of the light-matter coupling, $\sigma^{z}_{n}$ and $\sigma^{\pm}_{n}$ are the Pauli operators of the two-level atoms, $a_{n}$ $(a_{n}^{\dagger})$ is the annihilation (creation) operator of the cavity modes, $J$ is the photon hopping rate with a complex phase $e^{i\theta}$, $n=1,2,3$, and $\hbar\equiv1$. The complex phase can be engineered by applying an artificial vector potential $A(r)$ with $\theta=\intop_{r_{n}}^{r_{n'}}A(r)dr$ and $\theta\in [-\pi, \pi]$, which was realized using synthetic gauge fields~\cite{Roushan2017,Fan2019Sci,Baba2022SA, Oliver2024NP} or other interaction-engineering techniques~\cite{Bloch2016NC}.
The JC trimer possesses intrinsic U(1) symmetry because the total excitation number ${N}_{\mathrm{tot}}=\sum_{n}(a_{n}^{\dagger}a_{n}+\sigma^{+}_{n}\sigma^{-}_{n})$ is a conserved quantity given the commutation relation $[H,N_{\mathrm{tot}}]=0$. 

Below we study this JC trimer for a broad range of light-matter coupling strength while assuming weak photon hopping with $J/\omega_{c}\ll1$. Using a mean-field approach, we find solutions of the ground state phases in different parameter regimes. A phase diagram vs the dimensionless coupling strength $g_{1}=g/\sqrt{\omega_{0}\omega_{c}}$ and the phase $\theta$ is presented in Fig.~\ref{Fig1_model}(b).  
The system exhibits three distinctively different SPs in addition to the NP. 
We note that the JC model is usually obtained from the quantum Rabi model by taking the rotating-wave approximation to neglect the counterrotating terms in the limit of $g/\omega_{c}<0.1$. But for two-level atoms where the $\Delta m=\pm1$ transitions are coupled to circularly polarized optical fields~\cite{Crisp1991}, the JC model is the exact form of the Hamiltonian and the coupling strength can be comparable to or even exceed the qubit or cavity frequencies. In addition, a JC model with such couplings can also be obtained from the quantum Rabi model by applying frequency modulation on the qubits~\cite{Liu2024,Huang2020}.

\section{Normal phase}
We first calculate the energy spectrum of the JC trimer in the NP where the cavity has no excitation in the ground state by applying the Schrieffer-Wolff transformation $U_{_{\mathrm {SW}}}^{\mathrm{NP}}=\prod_n \exp[\frac{g_{1}}{\sqrt{\eta}}(a_{n}^{\dagger}\sigma^{-}_{n}-\sigma^{+}_{n}a_{n})]$ to each JC model with $\eta=\omega_{0}/\omega_{c}$~\cite{Wolff1966}. In the limit of $\eta\rightarrow\infty$ with $J\ll g, \omega_{0}$, the Hamiltonian to the second order of $1/\sqrt{\eta}$ can be written as $H_{_\mathrm{NP}}=\sum_n [\widetilde{H}_{_\mathrm{JC}}^{(n)} + H_{\mathrm{hop}}^{(n)} +\omega_c g_1^2/2]$, where the Hamiltonian of the JC model becomes $\widetilde{H}_{\mathrm{JC}}^{(n)} = \widetilde{\omega}_{0}\sigma^{z}_{n}/{2}+\omega_{c}a_{n}^{\dagger}a_{n} + 
g_{1}^{2}\omega_{c}a_{n}^{\dagger}a_{n}\sigma^{z}_{n}$ with $\widetilde{\omega}_{0} = \omega_{0}+\omega_{c}g_{1}^{2}$. 
With the atoms in the ground state $|\!\!\downarrow\rangle=\prod_{n}|g\rangle_{n}$, the Hamiltonian of the cavity modes is $H_{_\mathrm{NP}}^{\downarrow} = \langle\downarrow\!\!|H_{_\mathrm{NP}}|\!\!\downarrow\rangle$ with 
$H_{_\mathrm{NP}}^{\downarrow}= \sum_{n} [\widetilde{\omega}_{c}a_{n}^{\dagger}a_{n} +  H_{\mathrm{hop}}^{(n)} - \widetilde{\omega}_{0}/{2}]$ and $\widetilde{\omega}_{c} = \omega_{c}(1 -g_{1}^{2}) $. This Hamiltonian can be solved by performing a discrete Fourier transformation on the cavity operators: $a_{n}=\frac{1}{\sqrt{3}}\sum_q e^{- i n q}a_{q}$, where $a_{q}$ is the quasiparticle operator and 
$q=0, \pm2\pi/3$ is the quasimomentum. The Hamiltonian then becomes $H_{_\mathrm{NP}}^{\downarrow}= \sum_q \epsilon_{q}a_{q}^{\dagger}a_{q}+E_{g}^{^\mathrm{NP}}$ with the eigenenergy $\epsilon_{q}=\widetilde{\omega}_{c}+2J\cos(\theta-q)$ for mode $q$ and ground-state energy $E_{g}^{^\mathrm{NP}}\!\!=\!-3\widetilde{\omega}_{0}/2$. 
It can be shown that $\epsilon_{q}\geqslant0$ when the scaled coupling satisfies $g_{1}\leqslant g_{1c}$ with \begin{equation}
    g_{1c}=\sqrt{1+(2J/\omega_{c})\cos(\theta - q_\theta)}\
\end{equation}
being the critical coupling strength and $q_\theta=-\frac{2\pi}{3}, 0, \frac{2\pi}{3}$ for $\theta\in[0, \frac{2\pi}{3}]$,  
$\theta\in[\frac{2\pi}{3}, \frac{4\pi}{3}]$, and $\theta\in[-\frac{2\pi}{3},0]$, respectively. 
Hence, when $g_{1}<g_{1c}$, the ground state of this system is in the NP, and when $g_{1}>g_{1c}$, the NP becomes unstable and the system enters the SPs. The phase transition from NP to SP is second order with one quasiparticle mode $\epsilon_{q_\theta}=0$ at the critical point $g_{1c}$.  
The ground state of the NP in the physical frame can be written as $|G\rangle _{_\mathrm{NP}}=U_{_{\mathrm{SW}}}^{\mathrm{NP}} \left(\prod_q|0_q\rangle |\!\!\downarrow\rangle\right)$ with $|0_q\rangle$ being the vacuum state of the quasiparticle modes. As the unitary transformation $U_{_{\mathrm{SW}}}^{\mathrm{NP}}$ conserves the total excitation number, the ground state $|G\rangle _{_\mathrm{NP}}$ has zero photon with $\langle a_{n}\rangle =0$, and the atoms are in the ground state $|g\rangle$. We will use $\langle a_{n}\rangle/\sqrt{\eta}$ as the order parameter to distinguish different quantum phases in this system. Notably, the critical coupling $g_{1c}$ depends only on $J/\omega_c$ and $\theta$ and is independent of the frequency ratio $\eta$.

\section{Superradiant phases} 
In the SPs when $g_1>g_{1c}$, the cavity field is macroscopically occupied. Here we 
adopt a mean-field approach to solve the ground state of the SPs by replacing the cavity modes with $a_{n}\rightarrow b_{n}+\alpha_{n}$, where $\alpha_{n}=\langle a_{n} \rangle=A_{n}+iB_{n}$ is the complex cavity amplitude of the ground state, serving as the mean field in this model, and $b_n$ describes the quantum fluctuations around the average field amplitude. This approach has been widely used to study NP-SP phase transitions \cite{Emary2003,Hwang2015,Hwang2016,Hwang2022,He2024}. Higher-order terms of the fluctuation operator $b_n$ are omitted in this approach.
The total Hamiltonian $H$ then becomes $H_{_\mathrm{SP}} = \sum_n \left [H_{_\mathrm{QR}}^{(n)} + H_{\mathrm{hop}}^{(n)} + V_{1}^{(n)} \right] +C_1 $. Here
\begin{equation}
    H_{_\mathrm{QR}}^{(n)} = H_0^{(n)} + \frac{g^{2}}{\Delta_{n}}\left[\alpha_{n}b_{n}^{\dagger}(\Omega_{n}\text{\ensuremath{\tau}}^{+}_{n}-\Omega^{-1}_{n}\text{\ensuremath{\tau}}^{-}_{n})+\mathrm{H.c.}\right]
\end{equation}
is a quantum Rabi model that includes both rotating and counterrotating terms with $\tau^{z}_{n}=\frac{2gA_{n}}{\Delta_{n}}\sigma^{x}_{n}-\frac{2gB_{n}}{\Delta_{n}}\sigma^{y}_{n}+\frac{\omega_{0}}{{\Delta_{n}}}\sigma^{z}_{n}$ being the Pauli operator for the rotated atomic states, $H_0^{(n)}=\frac{\Delta_{n}}{2}\tau^{z}_{n}+\omega_{c}b_{n}^{\dagger}b_{n}$,
$\Delta_{n}=\sqrt{4g^{2}A_{n}^{2}+4g^{2}B_{n}^{2}+\omega_{0}^{2}}$ being the normalization factor, and
$\Omega_{n}=\sqrt{(\Delta_{n}-\omega_{0})/(\Delta_{n}+\omega_{0})}$. 
The term $H_{\mathrm{hop}}^{(n)}$ is now written in terms of the shifted cavity mode $b_n$.  
The term $V_{1}^{(n)}$ has the form  
\begin{eqnarray}
V_{1}^{(n)} &=& \biggl[\left(\omega_{c}+\frac{g^{2}}{\Delta_{n}}\text{\ensuremath{\tau}}_{n}^{z}\right)(\alpha_{n}b_{n}^{\dagger}+\alpha_{n}^{\ast}b_{n}) \nonumber \\
&&+J(e^{i\theta}b_{n}^{\dagger}\alpha_{n+1}+e^{-i\theta}b_{n+1}^{\dagger}\alpha_{n}+\mathrm{H.c.})\biggr]
\end{eqnarray} 
and the constant $C_{1}\!=\!\sum_{n}\left(\omega_{c}|\alpha_{n}|^{2}+2J{\rm Re}[e^{i\theta}\alpha_{n}^{\ast}\alpha_{n+1}]\right)$.  
To solve its ground state, we first apply the Schrieffer-Wolff transformation~\cite{Wolff1966} $U_{_{\mathrm{SW}}}^{\mathrm{SP}} = \prod_{n} \mathrm{exp} [\frac{g^{2}\alpha^{\ast}_{n}}{\Delta_{n}^{2}} (\Omega_{n}^{-1} \tau^{+}_{n}+\Omega_{n}\tau^{-}_{n})b_{n}-\mathrm{H.c.}]$ to the Hamiltonian $H_{_\mathrm{SP}}$, followed by projecting the Hamiltonian to the down-spin state $|\widetilde{\downarrow}\rangle=\prod_{n}|\widetilde{g}\rangle_{n}$ of the rotated spin operator $\tau^{z}_{n}$. 
The Hamiltonian $H_{_\mathrm{SP}}^{\widetilde{\downarrow}}=\langle \widetilde{\downarrow} | H_{_\mathrm{SP}} |\widetilde{\downarrow}\rangle$ reads
\begin{eqnarray}
    H_{_\mathrm{SP}}^{\widetilde{\downarrow}} &=&  \sum_{n=1}^{3}\bigg[\mu_{n} b_{n}^{\dagger}b_{n}\!+\!D_{n}b_{n}^{\dagger}+D_{n}^{*}b_{n}\nonumber\\
    & &+\left(\frac{\nu_{n}}{2}b_{n}^{\dagger2}\!+\!Je^{i\theta}b_{n}^{\dagger}b_{n+1}\!+\!\mathrm{H.c.}\right)\bigg]+C_{2}
    \label{HSP}
\end{eqnarray}
with $\mu_{n}=\omega_{c}-\frac{2g^{4}|\alpha_{n}|^{2}(\Delta_{n}^{2}+\omega_{0}^{2})}{\Delta_{n}^{3}(\Delta_{n}^{2}-\omega_{0}^{2})}$, $\nu_{n}=\frac{2g^{4}\alpha_{n}^{2}}{\Delta_{n}^{3}}$,
$D_{n}=(\omega_{c}\!-\!\frac{g^{2}}{\Delta_{n}})\alpha_n\!+\!J(e^{i\theta}\alpha_{n+1}+e^{-i\theta}\alpha_{n-1})$, and $C_{2} =C_{1}-\sum_{n}\left[\frac{\bigtriangleup_{n}}{2}-\frac{2g^{4}|\alpha_{n}|^{2}\omega_{0}}{\Delta_{n}^{2}(\Delta_{n}^{2}-\omega_{0}^{2})}\right]$.
If we use the mean-field approach,
the Hamiltonian $H_{_\mathrm{SP}}^{\widetilde{\downarrow}}$ should contain only quadrature terms with the linear terms vanishing, i.e., $D_{n}\equiv0$. Below we solve this equation in different parameter regimes.

(1) $Uniform$ $superradiant$ $phase$. When $|\theta|>\theta_{c}=2\pi/3$, the equation $D_{n}\equiv0$ yields an analytical solution with $\alpha_{n}\equiv\pm\frac{1}{2g}\sqrt{g^{4}/(\omega_{c}+2J\cos\theta)^{2}-\omega_{0}^{2}}$; i.e., the cavity displacements are real and uniform across the JC trimer. 
We denote $\alpha_n=A_0$, $\Delta_{n}=\Delta_{0}$, $\mu_{n}=\mu_0$ and $\nu_n = \nu_0$. Converting the Hamiltonian $H_{_\mathrm{SP}}^{\widetilde{\downarrow}}$ to the momentum space with $b_{n}=\frac{1}{\sqrt{3}}\sum_{q}e^{-i nq}b_{q}$, we obtain 
\begin{equation}    
H_{_{\mathrm{SP}}}^{\widetilde{\downarrow}}=\sum_{q}\left[\omega_{q}b_{q}^{\dagger}b_{q}+\frac{\nu_{0}}{2}\left(b_{q}^{\dagger}b_{-q}^{\dagger}+b_{q}b_{-q}\right)\right]+C_{2}
   \label{HSPF}
\end{equation}
with $\omega_{q}=\mu_0+2J\cos(\theta-q)$. This Hamiltonian is bilinear, including two-mode squeezing terms $b_{q}b_{-q}$ and $b_{q}^{\dagger}b_{-q}^{\dagger}$. The quasiparticle energy can be obtained as
\begin{equation}
	\epsilon_{q}=\frac{1}{2}(\omega_{q}-\omega_{-q})+\frac{1}{2}\sqrt{(\omega_{q}+\omega_{-q})^{2}-4\nu_0^2}.
\end{equation}
The ground state energy is $E_{g}^{\mathrm{USP}}=\frac{1}{2}\sum_{q}(\epsilon_{q}-\omega_{q})+C_{2}$.

(2) $Frustrated$ $superradiant$ $phases$. When $|\theta|<\theta_{c}$, the equation $D_{n}\equiv0$ can be solved numerically to derive the cavity displacements in the ground state. The displacements are nonuniform and complex, resulting in frustration in the SP.  
We define the operator vector $\vec{v}=(b_{1}^{\dagger},b_{2}^{\dagger}, b_{3}^{\dagger},b_{1},b_{2},b_{3})$. The Hamiltonian~(\ref{HSP}) is then $H_{_\mathrm{SP}}^{\widetilde{\downarrow}}= \frac{1}{2} \vec{v} M \vec{v}^{\dagger}+C_{3}$, where the Hermitian matrix 
\begin{equation}
M=\left(\begin{array}{cccccc}
\mu_{1} & Je^{i\theta} & Je^{-i\theta} & \nu_{1} & 0 & 0\\
Je^{-i\theta} & \mu_{2} & Je^{i\theta} & 0 & \nu_{2} & 0\\
Je^{i\theta} & Je^{-i\theta} & \mu_{3} & 0 & 0 & \nu_{3}\\
\nu_{1}^{\ast} & 0 & 0 & \mu_{1} & Je^{-i\theta} & Je^{i\theta}\\
0 & \nu_{2}^{\ast} & 0 & Je^{i\theta} & \mu_{2} & Je^{-i\theta}\\
0 & 0 & \nu_{2}^{\ast} & Je^{-i\theta} & Je^{i\theta} & \mu_{3}
\end{array}\right)
\end{equation}
with the constant $C_{3}=C_{2}-\frac{1}{2}\sum_{n}\mu_{n}$.
After the Hopfield-Bogoliubov transformation is appling~\cite{Hopfield1958}, the Hamiltonian can be diagonalized with $H_{_\mathrm{SP}}^{\widetilde{\downarrow}} = \sum_m \epsilon_{m}c_{m}^{\dagger}c_{m} + E_{g}^{\mathrm{FSP}}$, where $\epsilon_{m}$ is the quasiparticle energy with index $m$, $c_m$ ($c_m^{\dagger}$) is the annihilation (creation) operator of the quasiparticle mode and is a linear combination of the operators in $\vec{v}$, and $E_{g}^{\mathrm{FSP}}= C_{3} +\frac{1}{2}\sum_{m}\epsilon_{m}$ is the ground-state energy.
It can be shown that $E_{g}^{\mathrm{USP}}=E_{g}^{\mathrm{FSP}}$ at the boundaries $\theta=\pm\theta_{c}$.

\begin{figure}
    \center
    \includegraphics[clip, width=7cm]{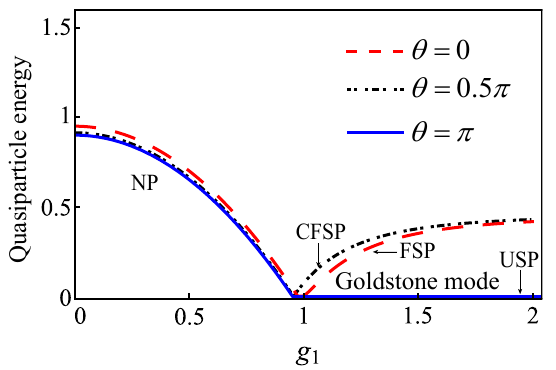}
    \caption{Quasiparticle energy vs $g_{1}$ for $\theta=0$, $0.5\pi$, and $\pi$. Here $\eta=10^3$, and $J/\omega_c=0.05$.}
    \label{Fig3_jifapu0.05}
\end{figure}
In Fig.~\ref{Fig3_jifapu0.05}, we plot the lowest quasiparticle energy $\epsilon_{q}$ (or $\epsilon_{m}$) vs the scaled coupling $g_{1}$ for several $\theta$ values. It can be observed that the quasiparticle energy approaches zero from both sides of the critical point $g_{1c}$, indicating a second-order phase transition at $g_{1c}$. 
For $\theta=0$ ($0.5\pi$), the phase transition from the NP to the FSP occurs at $g_1=g_{1c}=0.975$ ($0.956$) and $\epsilon_{m}$ increases with $g_1$ for $g_1>g_{1c}$. 
However for $\theta=\pi$, the transition from the NP to the USP occurs at $g_{1}=g_{1c}=0.949$. The quasiparticle energy remains at $\epsilon_{m}=0$ when $g_1>g_{1c}$, indicating a Goldstone mode in the USP induced by the spontaneous breaking of the U(1) symmetry~\cite{Goldstone1962,Hwang2016}.

\begin{figure}
    \center
    \includegraphics[clip, width=8cm]{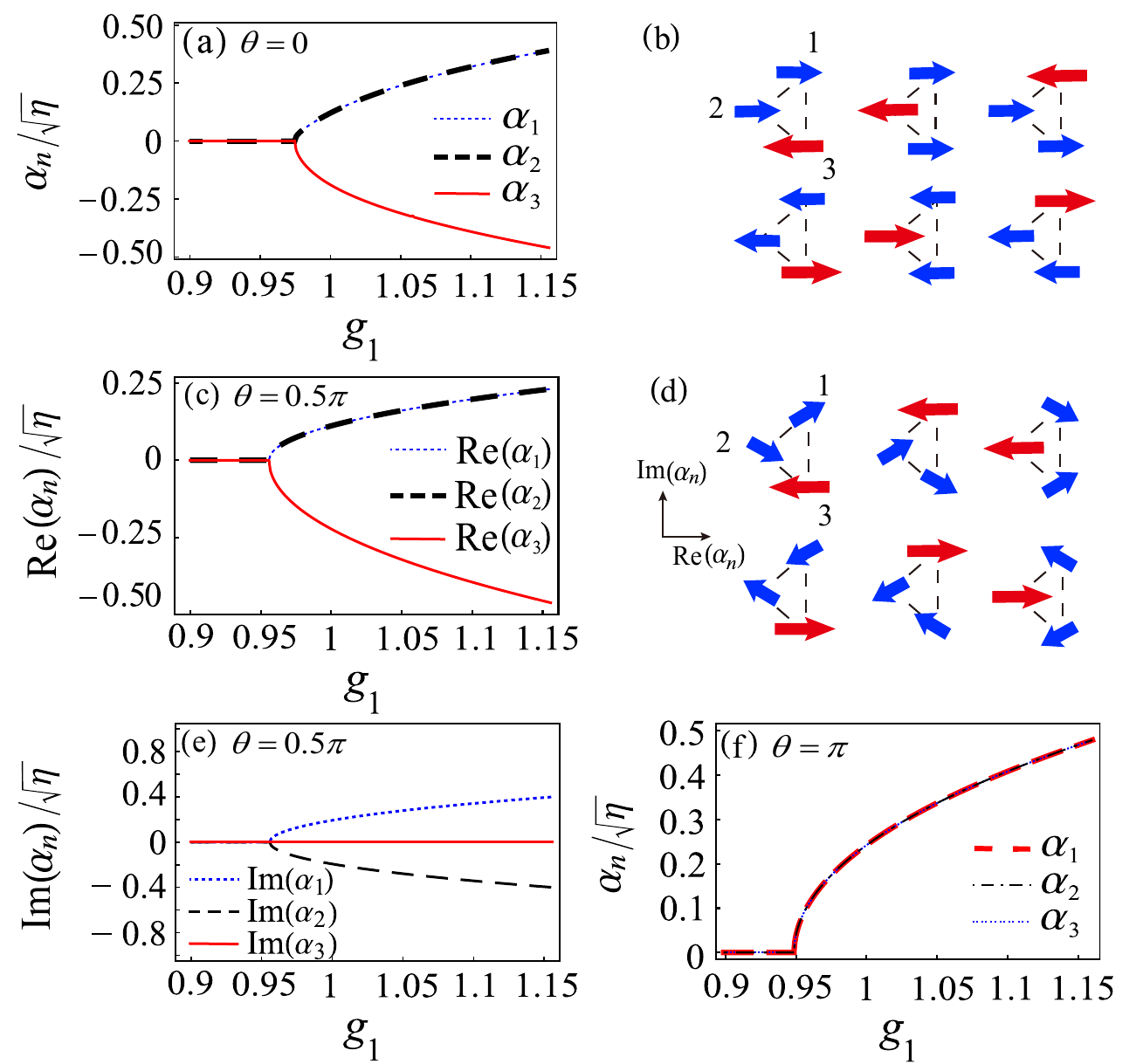}
    \caption{(a) Order parameter $\alpha_{n}/\sqrt{\eta}$ vs $g_{1}$ at $\theta=0$ and (b) its frustration configuration. (c) and (e) Real and imaginary parts of $\alpha_{n}/\sqrt{\eta}$ vs $g_1$ at $\theta=0.5\pi$ and (d) its frustration configuration. (f) Order parameter $\alpha_{n}/\sqrt{\eta}$ vs $g_{1}$ at $\theta=\pi$. Other parameters are the same as in Fig.~\ref{Fig3_jifapu0.05}.}
    \label{Fig3_pingyixvcanliang}
\end{figure}
\section{Quantum frustration}
For $|\theta|< 2\pi/3$ in the FSPs, the cavity displacements $\alpha_n$ are site dependent. At $\theta=0$, the displacements are real, and at $\theta\neq 0$, the displacements are complex. 
In Fig.~\ref{Fig3_pingyixvcanliang}(a), we plot one solution of $\alpha_{n}$ vs the scaled coupling $g_1$ at $\theta=0$. In the NP when $g_{1}<g_{1c}=0.975$, $\alpha_{n}=0$, while in the SP when $g_{1}>g_{1c}$, $\alpha_{n}\neq0$ with $\alpha_{1}=\alpha_{2}\neq\alpha_{3}$. The configurations of the cavity displacements are illustrated in Fig.~\ref{Fig3_pingyixvcanliang}(b), where the displacements are represented by blue and red arrows and the three displacements are in opposite directions to minimize the total energy. It can be shown that there are six different configurations, corresponding to six degenerate ground states. The displacements plotted in Fig.~\ref{Fig3_pingyixvcanliang}(a) correspond to one configuration in the first column of Fig.~\ref{Fig3_pingyixvcanliang}(b). 
At $\theta=0.5\pi$, the critical point is now $g_{1c}=0.956$ and the cavity displacements in the FSP are complex. Figures~\ref{Fig3_pingyixvcanliang}(c) and~\ref{Fig3_pingyixvcanliang}(e) plot the real and imaginary parts of one solution of $\alpha_{n}$, with $\alpha_{3}$ being real, $\alpha_{1,2}$ being complex, and $\alpha_{1}\neq\alpha_{2}\neq\alpha_{3}$. When $g_1> g_{1c}$, the system is in the CFSP with nontrivial quantum frustration and chirality. The configurations of the cavity displacements are given in Fig.~\ref{Fig3_pingyixvcanliang}(d), where the directions of the arrows indicate the real and imaginary components of the displacement and the chirality of the states.  
Therefore, the ground state of the JC trimer can be in the FSP's, demonstrating the quantum frustration phenomenon due to the triangular geometry of this system~\cite{Hwang2022,Furukawa2015,Fu2015,Diep2005,Lacroix2011}. 

For comparison, we plot the cavity displacements vs $g_1$ at $\theta=\pi>\theta_c$ in Fig.~\ref{Fig3_pingyixvcanliang}(f). Here, when $g_{1}>g_{1c}=0.949$, the ground state is in the USP with all cavities having equal displacements and no frustration.

We study the critical scaling behavior of the average photon number near the phase transition and numerically obtain
\begin{align}
& \langle a_{1}^{\dagger}a_{1}\rangle_{\rm FSP}\propto |g_{1}-g_{1c}|^{-1}, \nonumber\\
& \langle a_{2(3)}^{\dagger}a_{2(3)}\rangle_{\rm FSP}\propto |g_{1}-g_{1c}|^{-1/2}, \nonumber\\
& \langle a_{1}^{\dagger}a_{1}\rangle_{\rm CFSP}\propto |g_{1}-g_{1c}|^{-1/3}, \nonumber\\
& \langle a_{2(3)}^{\dagger}a_{2(3)}\rangle_{\rm CFSP}\propto |g_{1}-g_{1c}|^{-1/2}, \nonumber\\
& \langle a_{n}^{\dagger}a_{n}\rangle_{\rm USP}\propto |g_{1}-g_{1c}|^{-1/2}.
\end{align}
These critical exponents coincide with those reported for the quantum Rabi trimer~\cite{QuantumFuctuations2024}, indicating that the JC trimer and the quantum Rabi trimer belong to the same universality class despite the absence of counterrotating interactions in the former.

\begin{figure}
    \center
    \includegraphics[clip, width=8cm]{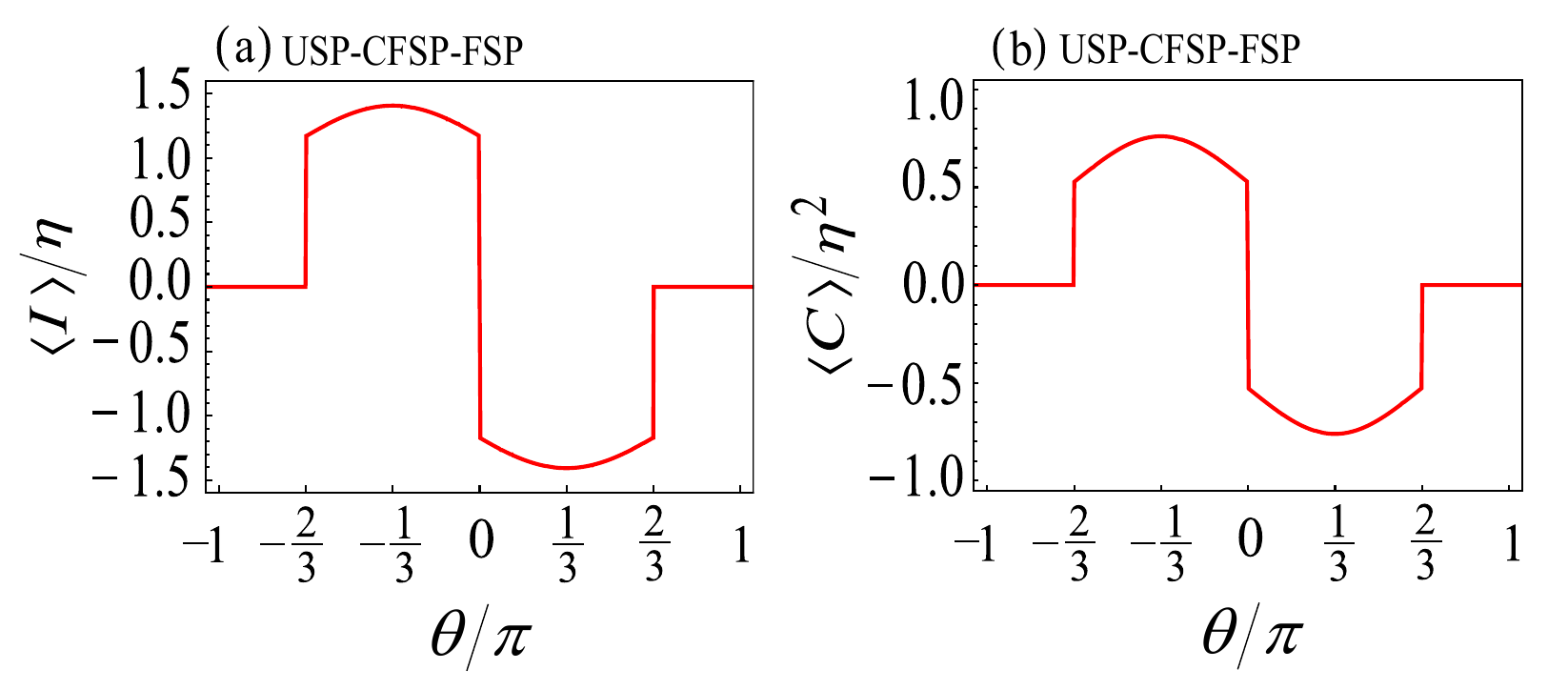}
    \caption{(a) Scaled photon current $\langle I\rangle/\eta$ vs the phase $\theta$. (b) Scaled chirality $\langle C\rangle/\eta^2$ vs $\theta$. Here $g_{1}=1.2>g_{1c}$, and other parameters are the same as in Fig.~\ref{Fig3_jifapu0.05}.}
    \label{Fig5_shouxing}
\end{figure}   
\section{Photon current and chirality} The cavity displacements in the CFSP ($0<\vert\theta\vert < \theta_c$) are complex, indicating nontrivial photon flow within the JC trimer. To characterize the photon flow, we define $I = i [(a_{1}^{\dagger}a_{2}+a_{2}^{\dagger}a_{3}+a_{3}^{\dagger}a_{1})-\mathrm{H.c.}]$ as the current operator~\cite{Roushan2017}. 
Figure~\ref{Fig5_shouxing}(a) plots the scaled photon current $\langle I\rangle/\eta$ vs the phase $\theta$ in the SPs with $g_1>g_{1c}$ for the entire range of $\theta$. It can be seen that $\langle I\rangle/\eta=0$ in the USP and the FSP but is nonzero in the CFSP. 
The abrupt change in $\langle I\rangle/\eta$ vs $\theta$ demonstrates a first-order phase transition between the three SPs. 

Nonzero photon flow in the ground state can be related to chiral symmetry breaking. To characterize the chirality of the SPs, we use the photon chiral operator $C= -2i \sum_{i,j,k} \varepsilon_{ijk} a_{i}a_{j}^{\dagger}(n_{k}-1/2)$, where $\varepsilon_{ijk}$ is the Levi-Civita tensor~\cite{Wen1989,Zhang2021}. The photon chiral operator satisfies $C_{r}^{-1}CC_{r}=-C$ under the chiral transformation $C_{r}$ that switches between even and odd permutations with $(123\leftrightarrow321)$, and it satisfies $T^{-1}CT=-C$ under the time-reversal transformation $T$. 
In Fig.~\ref{Fig5_shouxing}(b), we plot $\langle C\rangle/\eta^2$ vs $\theta$ in the SPs. For the USP and the FSP, $\langle C\rangle/\eta^2=0$. However, $\langle C\rangle>0$ at $-\theta_{c}<\theta<0$ and $\langle C\rangle<0$ at $0<\theta<\theta_{c}$ in the CFSP. This result indicates the broken chiral symmetry in the CFSP. The sign of $\langle C\rangle$ agrees with the sign of the current flow $\langle I\rangle$. The breaking of the chiral symmetry arises from time-reversal symmetry breaking introduced by the phase factor $e^{i\theta}$, which also induces a unidirectional photon current. 

\section{Discussion and Conclusions}
Strong light-matter coupling has been realized in a variety of experimental
platforms, including circuit-QED~\cite{Forn2019,Blais2021},
cavity-QED~\cite{Kimble1998}, and trapped-ion~\cite{cai} systems. In particular,
light-matter coupling strengths comparable to or even exceeding the cavity
frequency have been achieved in superconducting circuits~\cite{Yoshihara2017,Niemczyk2010} and semiconductor quantum wells~\cite{Bayer2017}. These developments provide a promising experimental foundation for exploring the superradiant phases predicted in this work. For atomic transitions with angular-momentum difference $\Delta m=\pm1$ coupled to a circularly polarized field, the JC interaction arises naturally without invoking the rotating-wave approximation, yielding the exact Hamiltonian $H_{\rm JC}$~\cite{Crisp1991}. Furthermore, an effective JC model with large normalized coupling strength can be engineered from the quantum Rabi model by suppressing the counterrotating interactions through frequency-modulation techniques~\cite{Huang2020,Liu2024}. The quantum phase transition studied here is characterized by the effective coupling $g_1=g/\sqrt{\omega_0\omega_c}$. The critical value $g_{1c}\approx$0.95$-$0.98 for $J/\omega_c=0.05$ depends on the hopping
phase $\theta$ but is independent of the frequency ratio $\eta=\omega_0/\omega_c$. Our theoretical analysis is performed in the large-$\eta$ regime. For finite $\eta$, the corresponding coupling ratios are $g/\omega_c=g_1\sqrt{\eta}$ and $g/\omega_0=g_1/\sqrt{\eta}$. Large frequency ratios and coupling strengths can be accessed in several quantum platforms~\cite{Yoshihara2017,Niemczyk2010,chen,cai,Zheng}, while frequency-modulation techniques provide an alternative route to engineering the required effective parameters
~\cite{Huang2020,Liu2024}. Furthermore, the complex hopping amplitudes required in our model can be engineered using synthetic gauge fields
~\cite{Roushan2017,Fan2019Sci,Baba2022SA,Oliver2024NP} and other
interaction-engineering techniques~\cite{Bloch2016NC}. Therefore, our
approach provides a possible route for exploring frustration and chirality
in light-matter systems.

The primary noise sources and decoherence mechanisms in the JC trimer are cavity photon loss, atomic relaxation, and dephasing. In the normal phase, the atoms remain close to their ground states and the cavity occupation is negligible, making dissipation effects relatively weak. In contrast, the superradiant phases are characterized by macroscopic cavity occupation and finite atomic excitation, so both the cavity and atomic dissipation can affect the stability and lifetime of the ordered phases. To observe the superradiant phases, the dissipation rates should remain small compared with the characteristic energy scale of the system, namely $\kappa/\omega_c \ll 1$ and $\gamma/\omega_c \ll 1$, where $\kappa$ and $\gamma$ denote the cavity and atomic dissipation rates, respectively. Such regimes were achieved in several cavity-~\cite{Zhang2014,Reagor2013} and circuit-QED platforms~\cite{Yoshihara2017,Niemczyk2010, Bosman2017,Oriani2025}. Therefore, the frustrated and chirally frustrated superradiant phases predicted here can persist over experimentally relevant timescales and can be detected through measurements of the cavity fields and photon currents.

To conclude, we investigated the superradiant phases of a JC trimer using a mean-field approach. In addition to the normal phase, the system exhibits three distinct superradiant phases: a uniform superradiant phase with uniform cavity displacements, a frustrated superradiant phase with nonuniform cavity displacements arising from geometric frustration, and a chirally frustrated superradiant phase characterized by broken chiral symmetry and unidirectional photon flow. We further characterized these phases through their excitation spectra, critical scaling behavior, photon currents, and chiral order parameters. Our results demonstrate that frustration, chirality, and superradiance can emerge in a minimal Jaynes-Cummings lattice without counterrotating interactions, providing a simple platform for exploring frustration-induced symmetry breaking and collective quantum phenomena in light-matter systems.


\begin{acknowledgments}
		  J.-F.H. is partially supported by the National Natural Science Foundation of China~(Grants No. 12075083 and No. 12475016) and Young Talents of the Furong Program of Hunan Province (Grant No. 2025RC3141).
\end{acknowledgments}

\appendix
\section{CounterRotating Terms}
To elucidate the connection between the JC trimer and the quantum Rabi triangle, we consider an anisotropic quantum Rabi trimer consisting of three identical cavity-QED systems arranged in a ring geometry. Each cavity is coupled to a two-level atom and is described by an anisotropic quantum Rabi model containing independent rotating and counterrotating interactions. Neighboring cavities are connected through photon hopping with a complex phase. The total Hamiltonian reads
\begin{eqnarray}
H & =&  \sum_{n=1}^{3}H_{\mathrm{AR}}^{(n)} +\sum_{n=1}^{3}J(e^{i\theta}a_{n}^{\dagger}a_{n+1}+\mathrm{H.c.}),\\
H_{\mathrm{AR}}^{(n)}& = &  \frac{\omega_{0}}{2}\sigma_{n}^{z}+\omega_{c}a_{n}^{\dagger}a_{n}+g_{r}(a_{n}^{\dagger}\sigma_{n}^{-}+a_{n}\sigma_{n}^{+})\nonumber \\
   &  & +g_{\mathrm{cr}}(a_{n}^{\dagger}\sigma_{n}^{+}+a_{n}\sigma_{n}^{-}).
\end{eqnarray}
Compared with the JC trimer Hamiltonian in Sec.~II, the light-matter interaction now contains both rotating and counterrotating terms, with coupling strengths $g_r$ and $g_{\mathrm{cr}}$, respectively. The model continuously interpolates between the JC trimer and the quantum Rabi triangle: When $g_{\mathrm{cr}}=0$, it reduces to the JC trimer studied in the main text, whereas for $g_{\mathrm{cr}}=g_r$, it becomes the quantum Rabi triangle considered in Ref.~\cite{Zhang2021}.

We introduce a Schrieffer-Wolff transformation of the anisotropic quantum Rabi trimer to eliminate the block-off-diagonal light-matter interaction. For site $n$ ($n=1,2,3$), the unitary operator of this transformation is
\begin{equation}
S_{\mathrm{np},n}=e^{-\left[\frac{g_{r}}{\omega_{0}}(a_{n}\sigma_{n}^{+}-a_{n}^{\dagger}\sigma_{n}^{-})+\frac{g_{\mathrm{cr}}}{\omega_{0}}(a_{n}^{\dagger}\sigma_{n}^{+}-a_{n}\sigma_{n}^{-})\right]}. 
\end{equation}
Applying $S_{\mathrm{np},n}$ to the interaction Hamiltonian $H_{\mathrm{AR}}^{(n)}$, we obtain
\begin{eqnarray}
H'_{\mathrm{AR},n} & = &  S_{\mathrm{np},n}^{\dagger}H_{\mathrm{AR},n}S_{\mathrm{np},n}\nonumber \\
 &  \approx  &  \omega_{c}a_{n}^{\dagger}a_{n}+\frac{\omega_{0}}{2}\sigma_{n}^{z}+\frac{g_{r}^{2}}{\omega_{0}}a_{n}^{\dagger}a_{n}\sigma_{n}^{z}+\frac{g_{\mathrm{cr}}^{2}}{\omega_{0}}a_{n}a_{n}^{\dagger}\sigma_{n}^{z}\nonumber  \\
 &  &  +\frac{g_{r}g_{\mathrm{cr}}}{\omega_{0}}\left(a_{n}^{\dagger2}+a_{n}^{2}\right)\sigma_{n}^{z}\nonumber  \\
 &  &
 +\frac{g_{r}^{2}}{\omega_{0}}\sigma_{n}^{+}\sigma_{n}^{-}-\frac{g_{\mathrm{cr}}^{2}}{\omega_{0}}\sigma_{n}^{+}\sigma_{n}^{-},
\end{eqnarray}
where higher-order corrections are neglected in the limit $J/\omega_0\ll1$ and $\eta=\omega_0/\omega_c\rightarrow\infty$.
Applying the transformation $U_{\mathrm{np}} = \Pi_{n=1}^{3}S_{\mathrm{np},n}$ on the full Hamiltonian of the trimer, we obtain
\begin{eqnarray}
H_{\mathrm{np}} & = & U_{\mathrm{np}}^{\dagger}HU_{\mathrm{np}}\nonumber \\
 & = & \sum_{n=1}^{3}H'_{\mathrm{AR},n}+J\sum_{n=1}^{3}(e^{i\theta}a_{n}^{\dagger}a_{n+1}+\mathrm{H.c.}).
\end{eqnarray}

Projecting the transformed Hamiltonian onto the spin-down subspace, we obtain an effective Hamiltonian for the normal phase,
\begin{eqnarray}
H_{\mathrm{np}}^{\downarrow} & = & \langle\downarrow|H_{\mathrm{np}}|\downarrow\rangle\nonumber \\
& =&  \sum_{n=1}^{3}[\omega_{c}(1-\lambda_{r}^{2}-\lambda_{\mathrm{cr}}^{2})a_{n}^{\dagger}a_{n}-\omega_{c}\lambda_{r}\lambda_{\mathrm{cr}}(a_{n}^{\dagger2}+a_{n}^{2})]\nonumber \\
& &  +J\sum_{n=1}^{3}(e^{i\theta}a_{n}^{\dagger}a_{n+1}+\mathrm{H.c.})-\frac{3\omega_{0}}{2}-3\omega_{c} \lambda_{\mathrm{cr}}^{2},
\end{eqnarray}
where the dimensionless coupling strengths are defined as $\lambda_{r}=g_{r}/ {\sqrt{\omega_{c}\omega_{0}}}$ and $\lambda_{\mathrm{cr}}=g_{\mathrm{cr}}/ {\sqrt{\omega_{c}\omega_{0}}}$. The effective Hamiltonian contains both photon-number-conserving terms and squeezing terms proportional to $\lambda_r\lambda_{\mathrm{cr}}$, which originate from the interplay between the rotating and counterrotating interactions.

To diagonalize the effective Hamiltonian, we perform a two-mode squeezing transformation of the form
\begin{equation}
S=\prod_{q}\exp\left[\frac{r_{q}}{2}\left(a_{q}^{\dagger}a_{-q}^{\dagger}-a_{q}a_{-q}\right)\right].
\end{equation}
Applying this transformation yields
\begin{eqnarray}
H_{\mathrm{np,eff}} & = &  S^{\dagger}H_{\mathrm{np}}^{\downarrow}S\nonumber \\
& = &  \sum_{q}[\omega_{1}a_{q}^{\dagger}a_{q}+\omega_{2}a_{-q}^{\dagger}a_{-q}+\omega_{3}(a_{q}a_{-q}+a_{q}^{\dagger}a_{-q}^{\dagger})]\nonumber \\
& & + \sum_{q}E_{q},
\end{eqnarray}
where
\begin{eqnarray}
\omega_{1} & =  &\omega_{q}\cosh^{2}r_{q}-2\omega_{c}\lambda_{r}\lambda_{\mathrm{cr}}\cosh r_{q}\sinh r_{q},\nonumber\\
\omega_{2} & =  &\omega_{q}\sinh^{2}r_{q}-2\omega_{c}\lambda_{r}\lambda_{\mathrm{cr}}\cosh r_{q}\sinh r_{q},\nonumber\\
\omega_{3} & = &  \omega_{q}\cosh r_{q}\sinh r_{q}-\omega_{c}\lambda_{r}\lambda_{\mathrm{cr}}\cosh2r_{q},\nonumber\\
E_{q} & =  &-2\omega_{c}\lambda_{r}\lambda_{\mathrm{cr}}\cosh r_{q}\sinh r_{q}+\frac{1}{2}\left(\omega_{q} +\omega_{-q}\right)\sinh^{2}r_{q} \nonumber \\
& &- \omega_{c}\lambda_{\mathrm{cr}}^{2}.
\end{eqnarray}
The squeezing parameter $r_q$ is chosen such that the off-diagonal terms proportional to $a_{q}a_{-q}+a_{q}^{\dagger}a_{-q}^{\dagger}$ vanish. This condition yields
\begin{eqnarray}
r_{q} & = & \frac{1}{4}\ln\frac{\omega_{q}+\omega_{-q}+4\omega_{c}\lambda_{r}\lambda_{\mathrm{cr}}}{\omega_{q}+\omega_{-q}-4\omega_{c}\lambda_{r}\lambda_{\mathrm{cr}}}.
\end{eqnarray}

The final form of the Hamiltonian for the normal phase is then
\begin{eqnarray}
H_{\mathrm{np,eff}} & = &   \sum_{q}\varepsilon_{q}a_{q}^{\dagger}a_{q}+E_{q},
\end{eqnarray}

where the excitation energy
\begin{equation}
\varepsilon_{q}  =  \frac{1}{2}[\sqrt{\left(\omega_{q}+\omega_{-q}\right)^{2}-16\omega_{c}^{2}\lambda_{r}^{2}\lambda_{\mathrm{cr}}^{2}}+\omega_{q}-\omega_{-q}],    
\end{equation}
and the ground state energy 
\begin{equation}
    E_{q}  =  \sum_{n=1}^{3}(\frac{1}{2}\varepsilon_{q}-\frac{1}{2}\omega_{q}-\omega_{c}\lambda_{\mathrm{cr}}^{2}).
\end{equation}
\begin{figure}
\includegraphics[clip, width=8.5cm]{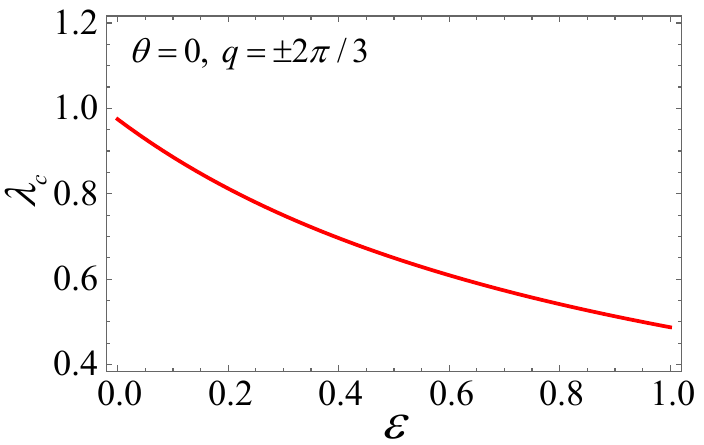}
\caption{The critical coupling strength $\lambda_{c}$ vs the anisotropy parameter $\varepsilon$ for $J/\omega_{c}=0.05$.} 
\label{fig5}
\end{figure}
The phase boundary between the normal phase and the superradiant phase is determined by the condition that the quasiparticle excitation gap closes,
\begin{equation}
\varepsilon_{q} = 0.
\end{equation}
To investigate the evolution from the JC trimer to the quantum Rabi triangle, we parametrize the coupling strengths as $\lambda_{r}\equiv\lambda$ and $\lambda_{\mathrm{cr}}\equiv\varepsilon\lambda$, where the anisotropy parameter $\varepsilon$ characterizes the relative strength of the counterrotating interaction. The model reduces to the JC trimer at $\varepsilon=0$ and continuously evolves into the quantum Rabi triangle at $\varepsilon=1$.

For $\varepsilon\neq1$, the critical coupling strength can be obtained analytically as
\begin{equation}
\lambda_{c}=\sqrt{\frac{A-\sqrt{B}}{\left(1-\varepsilon^{2}\right)^{2}}},
\end{equation}
where
\begin{eqnarray}
A & =&  \left(1+\varepsilon^{2}\right)\left(1+\frac{2J}{\omega_{c}}\cos\theta\cos q\right), \\
B & = &  4\varepsilon^{2}\left(1+\frac{2J}{\omega_{c}}\cos\theta\cos q\right)^{2} \nonumber \\
& &+\left(1-\varepsilon^{2}\right)^{2}\frac{4J^{2}}{\omega_{c}^{2}}\sin^{2}\theta\sin^{2}q.
\end{eqnarray}
The critical coupling depends explicitly on the anisotropy parameter $\varepsilon$, the hopping phase $\theta$, and the quasimomentum $q$, thereby providing a continuous description of the phase boundary from the JC limit ($\varepsilon=0$) to the quantum Rabi limit ($\varepsilon=1$).
Figure~\ref{fig5} plots the critical coupling strength $\lambda_{c}$ as a function of the anisotropy parameter $\varepsilon$. As $\varepsilon$ increases from 0 to 1, the critical coupling decreases monotonically from approximately $1$ to $0.5$. This behavior indicates that the counterrotating interaction facilitates the onset of the superradiant phase transition by reducing the rotating-wave coupling strength required to reach the critical point.


%
	
	
\end{document}